\documentclass[twocolumn, pra,showpacs]{revtex4}

	\usepackage{amsmath}
	\usepackage{subfigure}
	\usepackage{epsfig}
	\usepackage[colorlinks,hyperindex]{hyperref}
	\hypersetup
	{
		colorlinks,%
		citecolor=black,%
		linkcolor=black,%
		urlcolor=black,%
	}

\makeindex

\begin{document}

\title{ An Observed-Data-Consistent Approach to the Assignment of Bit Values in a Quantum Random Number Generator  }


\author{Pavel Lougovski} 
\email{lougovskip@ornl.gov}
\author{Raphael Pooser}

\affiliation{Quantum Information Science Group, Oak Ridge National Laboratory, Oak Ridge, TN 37831}


\begin{abstract}
 The majority of Quantum Random Number Generators (QRNG) are designed as converters of a continuous quantum random variable into a discrete classical random bit value. For the resulting random bit sequence to be minimally biased, the conversion process demands an experimenter to fully characterize the underlying quantum system and implement parameter estimation routines. Here we show that conventional approaches to parameter estimation (such as e.g. {\it Maximum Likelihood Estimation}) used on a finite QRNG data sample without caution may introduce binning bias and lead to overestimation of the randomness of the QRNG output. To bypass these complications, we develop an alternative conversion approach based on the Bayesian statistical inference method. We illustrate our approach using experimental data from a time-of-arrival QRNG and numerically simulated data from a vacuum homodyning QRNG. Side-by-side comparison with the conventional conversion technique shows that our method provides an automatic on-line bias control and naturally bounds the best achievable QRNG bit rate for a given measurement record. 
\end{abstract}



\maketitle

\section{Introduction}
Random numbers are important for an array of applications from encryption and authentication systems \cite{encryption}, to Monte Carlo simulations for molecular dynamics, nuclear reactors, and others \cite{primality}.  As a result, a variety of classical methods (computational pseudo-random number generators, sampling stochastic physical processes, etc.) to generate random number sequences have been developed. An attendant host of tests to certify that a given data sequence is ``random'' has been also been created \cite{nistsuite,diehard,ent}. While pseudo-random numbers are useful for many of these applications, including simulations and encryption with suitably high quality sources, their inherent determinism means that any encryption or authentication scheme is in principle breakable with sufficient computational power. This principle applies to any deterministic system, including processes described by classical physics.

On the other hand, the only nondeterministic physical theory with experimentally accessible applications is quantum mechanics \cite{bustard}. The additional security provided by non determinism is a requirement for quantum key distribution, for instance, whose security proofs often rely on the concept of true, nondeterministic randomness in order to guarantee successful secret key sharing \cite{gisin}. Thus, a wide array of so-called quantum random number generators (QRNG) have been developed. From radioactive decay \cite{schmidt} to quantum optical techniques \cite{jennewein,stefanov}, a host of methods involving photon arrival time \cite{rogina,wayne,shields,Wahl} and vacuum noise measurements \cite{lam,Gabriel} have been demonstrated. Despite the prevalence of QRNGs and their acknowledged need, many implementations use extractors (such as hashes) to remove large amounts of bias computationally, exposing a potential weakness in their physical implementations. For instance, if an adversary is able to computationally reverse the extractor function that a given QRNG implements in order to achieve random number uniformity and the underlying (``physical'') distribution is strongly biased then he or she will have a best-guess strategy against the QRNG device. Therefore, one's ability to detect and remove bias before applying an extractor function improves the QRNG's security.

One of the major sources of bias in QRNGs, aside from environmental noise, is the lack of knowledge of precise values of the QRNG's physical parameters. The best one may do is to estimate the parameters statistically. But because the estimates are statistical they are intrinsically noisy, and thus assigning a single value to a parameter can lead to errors and bias. Nevertheless, parameter estimator errors are usually ignored in QRNG design and simple point estimators are used. Here, we show that using point estimators may introduce possible binning bias. We argue that using a Bayesian statistical inference method removes this type of bias and propose a binning scheme that extracts the optimum number of bits possible for a given entropy from a given physical random number distribution. When used as a diagnostic for QRNGs in combination with maximum likelihood estimators (MLE), uniform distributions can be generated from sources of quantum randomness. Using Bayesian hypothesis updating techniques, our scheme allows for a test of the quantum model that produced a given set of numbers, potentially allowing for a fast, on-line quantum test of randomness. This technique has applications to high bit rate QRNGs which need testing and verification to ensure the device remains bias-free during use.
\vspace*{-0.71cm}\section{Direct Binning from a Continuous Distribution and Bias}\label{SEC:DirectBinning}
\par
Let $X$ be a continuous random variable with probability density function ({\it pdf}) $f_{X}(x|\theta)$, where $\theta$ is a fixed- (but unknown-) value parameter. The particular form and parametric dependence of $f_{X}(x|\theta)$ is determined by the experimental setup at hand. Our goal in this section is to introduce a typical problem of physical random number generation that can be formulated as follows: Provided $M$ independent samples of $X$, $\{x_{1},\dots,x_{M}\}$ are measured in an experiment, convert, if possible, each measurement outcome $x_{i}, i=1,M$ into a discrete random variable $K=k_{i}$ with the  probability mass function ({\it pmf}) $f_K(k)$ and corresponding  domain $\mathcal{K}=\{1,\dots,N\}$. A uniform distribution $U(1,N)$ is often important in applications and here we will also concentrate on the case of $f_K(k)=U(1,N)$. Then the problem essentially reduces to constructing a surjection $\mu$ from the set $\mathcal{X} = \{x:f_{X}(x|\theta)> 0\}$ onto the set $\mathcal{K}=\{k:f_{K}(k)=1/N,k=1,\dots,N\}$. 
\begin{figure}[tl]
	\begin{center}
		\includegraphics[scale = 0.4]{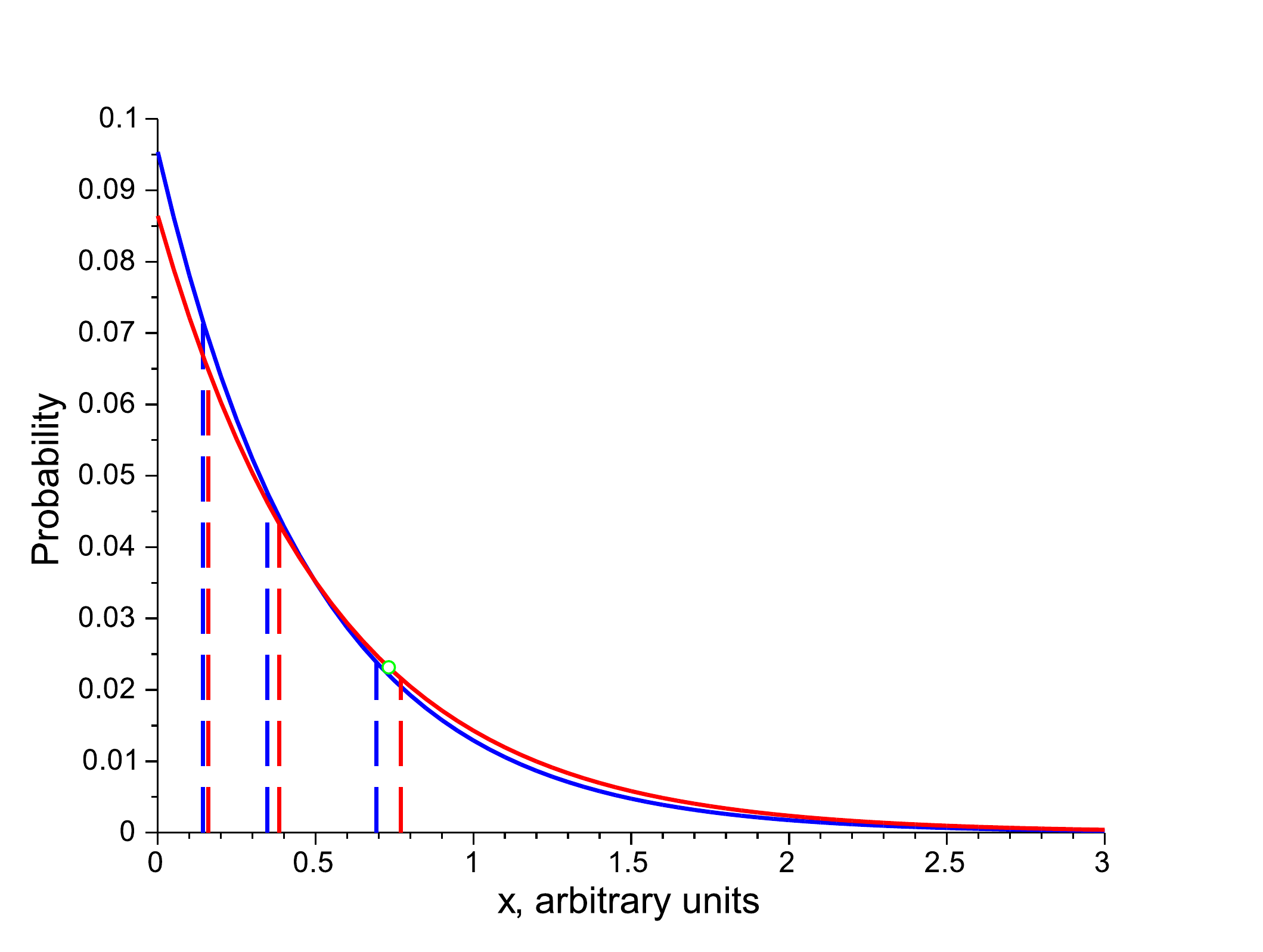}
		\caption{Example of binning ambiguity for different values of distribution parameter.}
		\label{Fig:BinningExample1}
	\end{center}
\end{figure}

Traditionally, the problem is solved by dividing the domain of $f_{X}(x|\theta) $, $\mathcal{X}$, into $N$ mutually non-intersecting bins such that $\mathcal{X} = \{B_{1}\cup \cdots \cup B_{N}\}$\cite{Qi}. When bins are selected such that the probability of the random variable  $X$ to fall into the $i$-th bin $B_{i}$ is
\begin{equation}\label{Eq:BinProbability1}
	P(X\in B_{i}|\theta) = \int_{B_{i}} dx f_{X}(x|\theta) = \frac{1}{N}, \forall i, 
\end{equation}
 then the surjection $\mu : \mathcal{X}\rightarrow \mathcal{K}$ can be constructed by following a simple rule: If a measurement result $X=x^{\prime}\in B_{i}$ for some $i$ then we assign $K=i$. Of course this mapping works only if the value of the model parameter $\theta$ is known. Since it is usually not the case in the majority of experimental situations, the first order of business is to find a good estimate of the value of $\theta$. In many cases, the number of possible ways to construct an estimator that provides an unbiased estimate of $\theta$ is infinite \cite{CasellaBerger}. Moreover, it is not always possible to find an estimator that has minimal uncertainty, and often one is forced to choose one from a set of almost optimal candidates.  In practice the maximum likelihood estimator (MLE) is a common choice. 
 
Given a set $d$ of independent samples of $X$, $d=\{x_{1},\dots,x_{M}\}$, we can introduce the {\it likelihood} function,
 \begin{equation}
 L(\theta|d) = \prod_{k=1}^{M} f_{X}(x_{k}|\theta). 
\end{equation}
The likelihood $L(\theta|d)$ indicates which values of $\theta$ are more likely given measurement data $d$. We can also compute, at least numerically, the value of $\theta$ that maximizes $L(\theta|d)$, provided the likelihood function is convex. The resulting estimator is MLE, i.e. $\theta_{MLE}={\rm max}_{\theta} L(\theta|d)$.

Using $\theta_{MLE}$ as the ``true'' parameter value for binning purposes in Eq.(\ref{Eq:BinProbability1}) might at first appear as a reasonable choice, and this approach is a mainstay in QRNG design. But what happens if instead of $\theta_{MLE}$ one uses some other estimate $\theta^{\prime}$ that differs from $\theta_{MLE}$ only slightly in the value of the likelihood, i.e. $|L(\theta^{\prime}|d)-L(\theta_{MLE}|d)|\ll 1, \theta^{\prime}\neq \theta_{MLE}$? Choosing $\theta^{\prime}$ over $\theta_{MLE}$ will have an effect on the size of bins $B_{i}$ generated via Eq.(\ref{Eq:BinProbability1}). We illustrate this situation in Fig.~\ref{Fig:BinningExample1}, where the random variable $X$ follows gamma distribution $\Gamma(1,\theta)$ (i.e. $f_{X}(x|\theta) = \theta e^{-\theta x}$) and we are interested in converting each measurement outcome $X=x_{i}$ into a uniformly distributed discrete random variable K that can take on values \{0,1,2,3\}. We fit the same measurement data using two slightly different values of the parameter $\theta$. The red solid line represents the fit with $\theta_{1} = 1.8$ and the blue solid line has $\theta_{2} = 2.0$. The vertical dashed blue (red) lines represent bins calculated using Eq.(\ref{Eq:BinProbability}) with $N=4$ and $\theta = \theta_{2}$($\theta = \theta_{1}$). The green circle is a particular measurement outcome $x_{i}$ that we would like to assign a discrete value $k$ to. According to our previous discussion, $k = 3$ and $k = 2$ if we use values $\theta_{2}$  and $\theta_{1}$ respectively.

\begin{figure}[t]
	\begin{center}
		\includegraphics[scale = 0.4]{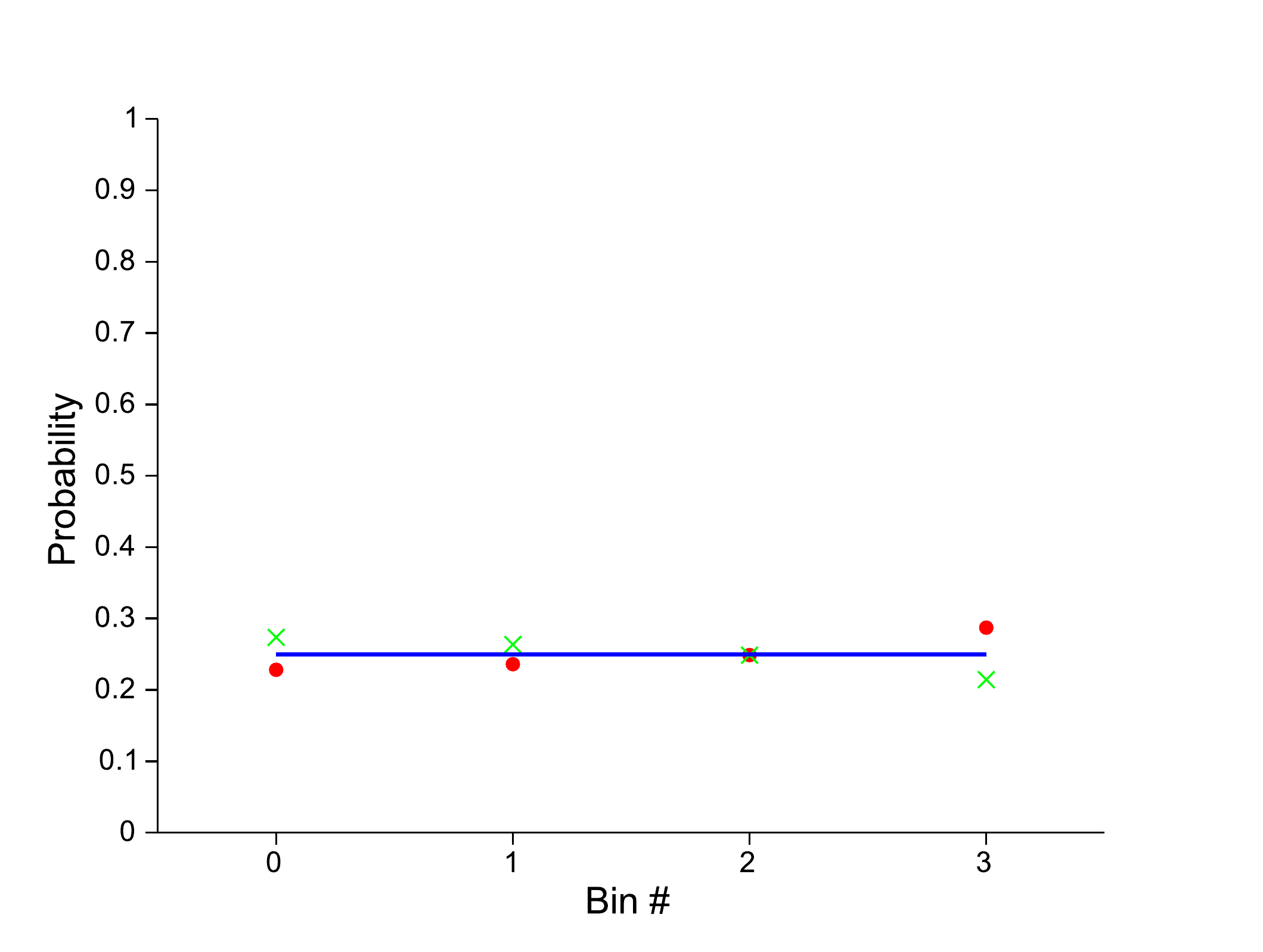}
		\caption{Example of bias due to parameter estimation uncertainty.}
		\label{Fig:BinProbability}
	\end{center}
\end{figure}

Now imagine that $\theta_{1}$ and $\theta_{2}$ are such that the likelihood function does not provide a reliable differentiation between them, i.e. $L(\theta_{1}|d)\approx L(\theta_{2}|d)$. Which value of $k$, if any,  should we then adopt? There are four possible options:
\begin{itemize}
\item Choose $\theta = \theta_{1}$ when $\theta_{1}$ is the true estimate ($k=2$).
\item Choose $\theta = \theta_{2}$ when $\theta_{2}$ is the true estimate ($k=3$).
\item Choose $\theta = \theta_{1}$ when $\theta_{2}$ is the true estimate ($k=2$).
\item Choose $\theta = \theta_{2}$ when $\theta_{1}$ is the true estimate ($k=3$).
\end{itemize} 
The first two choices are trivial since they obviously result in a uniform pmf  $f_{K}(k)=\frac{1}{4}\, \forall k$. The last two choices, however, generate a bias that distorts the uniformity of $f_{K}(k)$. To see that we calculate the probability of $X$ occupying the $i$-th bin provided that $\theta = \theta_{1}$ is chosen when $\theta_{2}$  is the true estimate,
\begin{align}
\nonumber P(X\in B_{i}|\{\theta = \theta_{1}|\theta_{2}\} ) & = \\
\nonumber P(X\in B_{i}|\theta = \theta_{1} {\rm when\,\theta_{2}\,is\,the\,true\,estimate}) & =  \\ 
  \int_{x_{i}}^{x_{i+1}} dx f_{X}(x|\theta_{1}) = [\frac{N-i}{N}]^{\frac{\theta_{1}}{\theta_{2}}} - [\frac{N-i-1}{N}]^{\frac{\theta_{1}}{\theta_{2}}},\label{Eq:BinProbUnder}
\end{align}
where $N=4$, $x_{i} = -\frac{1}{\theta_{2}}\ln(\frac{N-i}{N})$, and $i=0,1,2,3$. We notice that, by definition, $f_{K}(k=i) = P(X\in B_{i}|\{\theta = \theta_{1}|\theta_{2}\})$. Similarly, if   we choose $\theta = \theta_{2}$ when $\theta_{1}$ is the true estimate then the $i$-th bin probability $\tilde{f}_{K}(k=i)$ reads,
\begin{align}
\nonumber P(X\in B_{i}|\{\theta = \theta_{2}|\theta_{1}\} ) = \tilde{f}_{K}(k=i)& = \\
\nonumber P(X\in B_{i}|\theta = \theta_{2} {\rm when\,\theta_{1}\,is\,the\,true\,estimate}) & =  \\ 
  \int_{x_{i}}^{x_{i+1}} dx f_{X}(x|\theta_{2}) = [\frac{N-i}{N}]^{\frac{\theta_{2}}{\theta_{1}}} - [\frac{N-i-1}{N}]^{\frac{\theta_{2}}{\theta_{1}}}, \label{Eq:BinProbOver}
\end{align}
where   $x_{i} = -\frac{1}{\theta_{1}}\ln(\frac{N-i}{N})$. Finally, the plot of pmfs  $f_{K}(k)$ and $\tilde{f}_{K}(k)$ in Fig.~\ref{Fig:BinProbability}, calculated using Eqs.(\ref{Eq:BinProbUnder}) and (\ref{Eq:BinProbOver}) respectively for $N=4$, illustrates the effect of parameter under(over)-estimation on the uniformity of the random numbers generated using the continuous distribution binning method. The horizontal axis represents the bin number $k$ where a measurement outcome $x_{i}$ is placed as the result of binning. The vertical axis is the probability for different values of $k$ to occur. Ideally, if the value of $\theta$ was known exactly, the probability of $k=0,1,2,$ or $3$ would be the same at $\frac{1}{4}$. This situation is represented by the solid blue line. When the value of $\theta$ is overestimated, $\tilde{f}_{K}(k)$ -- the corresponding  pmf in Eq.(\ref{Eq:BinProbOver}) -- depicted by green crosses, exhibits a bias towards placing measurement outcomes into the first two bins. In a similar fashion, $f_{K}(k)$ in Eq.(\ref{Eq:BinProbUnder}), represented by red circles, corresponds to the situation when the parameter $\theta$ is underestimated and demonstrates bias towards $k=3$. To quantify the amount of introduced bias we compute values of Kullback-Leibler (KL) divergence $D_{KL}(\tilde{f}_{K}(k)||U(1,4))$ and $D_{KL}(f_{K}(k)||U(1,4))$ between the bin pmf $\tilde{f}_{K}(k)$ ($f_{K}(k)$) and the ideal uniform pmf $U(1,4) = \frac{1}{4}$ respectively. By definition, KL divergence measures the information lost when the uniform pmf $U(1,4)$ is used to approximate $\tilde{f}_{K}(k)$ or $f_{K}(k)$.  We find that $D_{KL}(\tilde{f}_{K}(k)||U(1,4)) = 0.006$ bits and $D_{KL}(f_{K}(k)||U(1,4)) = 0.0059$ bits.

This example shows that discrete random number generation procedures relying on binning a continuous probability distribution with a parametric dependence potentially introduces bias. This happens because the point parameter estimation approach is prone to over(under)-estimating the true value of the parameter. Hence, the question arises: Is there a binning method that does not introduce bias? The short answer is yes, and such a method will be introduced in Section~\ref{Sec:BayesianInference}. In the next section, a slightly different approach to binning is shown in order to motivate the discussion.

\section{Uniform Random Numbers via Integral Transform}
A measurement outcome $X=x_{i}$ does not depend on the value of the pdf parameter $\theta$. However, the probability of the outcome does. As we have already seen, this means that the size of the bins also depends on $\theta$, which makes binning procedure problematic. The reverse situation would be more practical, in which the bin size is fixed (independent of $\theta$) but the measurement outcome depends on the  pdf parameter. Of course, this does not remedy the problem of bias discussed earlier, but it will be useful in formulating a solution in the next section.

For a given fixed value $\theta$, the probability $P(X\le x|\theta)$ that the continuous random variable $X$ is less than $x$ reads, 
\begin{equation}\label{EQ:IntTrans}
P(X\le x|\theta) = \int_{-\infty}^{x} dt f_{X}(t|\theta)
\end{equation}
where we have assumed that  $X \in (-\infty,+\infty)$. By definition $P(X\le x|\theta)\in [0,1]$ and $U=U(x)=P(X\le x|\theta)$ can be interpreted as a uniform continuous random variable on the $[0,1]$ interval provided $P(X\le x|\theta)$ is a continuous function of $x$. The proof is straightforward and can be found elsewhere~\cite{CasellaBerger}. 
On the other hand, if the value of $x$ is fixed, e.g. $x=x^{\prime}$, and the value of $\theta$ is unknown then $U=P(X\le x^\prime | \theta) = U(\theta|x^\prime)$ is clearly a function of $\theta$ with the range $[0,1]$. 

If we divide the $[0,1]$ interval into $N$ uniform bins, each of the size $1/N$, then for every measurement outcome $X=x_{i}$ a discrete random number $K=k, k=\{1,\cdots,N\}$ can be generated by finding $k$ such that $ (k-1)/N \le U(\theta|x_{i}) < k/N$. This is exactly what we were looking for. By replacing the random variable $X$ with $U$ using the integral transform in Eq.(\ref{EQ:IntTrans}) we switched from having bins that explicitly depended on the model parameter $\theta$ to having constant bin size. The parametric dependence is now shifted to the random variable that we bin, i.e. $U(\theta)$, and now we need to figure out a way to assign a value to $U(\theta)$ which does not create bias.

\section{Bayesian Inference and Binning-Bias-Free Random Numbers}\label{Sec:BayesianInference}
We could try to fix the value of $U(\theta)$ by using an estimate of $\theta$ (e.g. MLE) as was done previously in Section~\ref{SEC:DirectBinning}. However, this approach is inherently flawed because any finite data sample estimator -- though it can be very close to the true parameter value -- will over(under)-estimate the true parameter value. However, the concept of likelihood, or, more precisely, the concept of treating the distribution parameter $\theta$ as an unknown (but not random) variable given a set of measurements  $d=\{x_{1},\dots,x_{M}\}$ can be inverted using Bayesian inference to compute the probability $U(\theta)$ of occupying a given bin $i$.

Indeed, the Bayesian approach treats $\theta$ as a quantity whose variation is described by a probability distribution $\pi(\theta)$ usually referred to as the {\it prior}. The prior is a subjective distribution determined by experimenter's personal beliefs and knowledge about the system of interest prior to any observations on the system.  Once $\pi(\theta)$ is formulated, an observation on the system is made. The prior is then updated with the result of the observation using Bayes rule and the next measurement is taken with the updated prior, often called {\it posterior}, as the new prior. If the sampling distribution, i.e. the distribution we draw measurement outcomes from,  is $f_{X}(x|\theta)$ (the pdf to observe $X=x$ as a result of our measurement, given the parameter value $\theta$) and the measurement result is $X=x$ then the posterior distribution is given by
\begin{equation}\label{Eq:BayesRule}
\pi(\theta|x) = \frac{f_{X}(x|\theta)\pi (\theta)}{m(x)},
\end{equation}
where $m(x)$ is the marginal distribution of $X$:
\begin{equation}
m(x) = \int d\theta f_{X}(x|\theta)\pi (\theta).
\end{equation}
The posterior distribution can be subjectively interpreted (since it does depend on the choice of the prior) as a conditional distribution (conditioned on the observed sample) for the parameter $\theta$. On the other hand, we know that $U(\theta)$  is a function of $\theta$ given the measurement outcome $X=x$. Therefore, $U(\theta)$ can also be interpreted as a random variable on $[0,1]$ with distribution function $g_{U}(u|x)$ that can be computed  using $\pi(\theta|x)$,
\begin{equation}\label{Eq:GofU}
g_{U}(u|x) = \pm \pi(U^{-1}(u)|x)\frac{d U^{-1}(u)}{d u},
\end{equation}
where the plus (minus) sign is taken when $U(\theta)$ is an increasing (decreasing) function of $\theta$, $U$ is assumed to be continuous, and $U^{-1}$ has a continuous first derivative. Now we are fully equipped to calculate the probability that a measurement outcome $X=x_{i}$ converts into an integer $k$ ($k\in[1,N]$). It is equivalent to the probability that the  random variable $U(\theta|x_{i})$ falls into the interval $ [(k-1)/N, k/N)$ given by,
\begin{align}\label{Eq:BinProbability}
\nonumber P(x_{i}\rightarrow k) & = P\left( \frac{k-1}{N}\le U(\theta|x_{i}) < \frac{k}{N} \right) = \\
& = \int_{ \frac{k-1}{N}}^{ \frac{k}{N}}  g_{U}(u|x_{i}) du.
\end{align}
This means that we now can assign a bin to a measurement outcome using a simple acceptance/rejection test: We accept $x_{i}$ into the $k$-th bin if $P(x_{i}\rightarrow k)\ge P_{a}$  and reject $x_{i}$ in the $k$-th bin otherwise. Here $P_{a}$ is the user-defined acceptance probability. The binning bias can be completely eliminated by setting the value of $P_{a}$ high ($P_{a}\ge 0.95$). This means that only the measurement outcomes that have more than 95\% of their distribution function $g_{U}(u|x_{i})$ localized within a certain bin will be accepted and converted into a discrete random number. All other measurements will be rejected. On the other hand, if $P_{a}$ is set too low, say, $P_{a}<0.5$ then less measurements will be rejected. However, this may lead to conflicting situations when a measurement outcome could be placed into two or more different bins which, in turn, may lead to binning bias. 
\begin{figure}[t]
	\begin{center}
		\includegraphics[scale = 0.4]{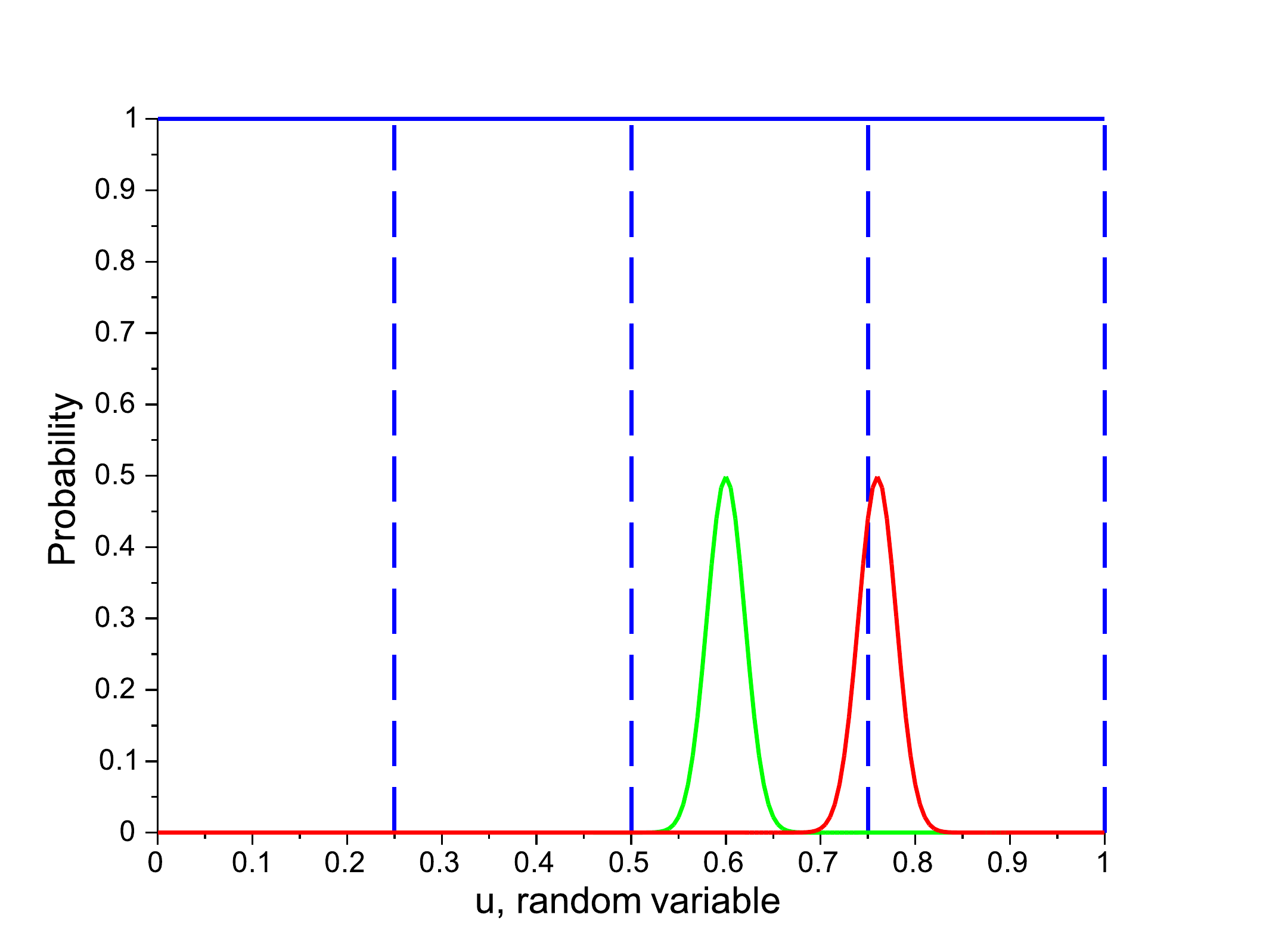}
		\caption{Example of measurement acceptance/rejection based on $U(\theta)$ probability distribution.}
		\label{Fig:BinVsParam}
	\end{center}
\end{figure}

Let us consider an example depicted in Figure~\ref{Fig:BinVsParam} where two distribution functions $g_{U}(u|x_{1})$ (red solid line) and $g_{U}(u|x_{N})$ (green solid line) for two independent samples $x_{1}$ and $x_{N}$ are plotted. We are interested in converting each measurement outcome $x$ into an integer value $\{0,1,2,3\}$. Using our acceptance/rejection test with $P_{a}=0.95$ we conclude that $x_{N}$ is an acceptable measurement that can be converted to $k = 2$. On contrary, $x_{N}$ will be rejected and no integer value will be assigned to it.    

We finally summarize our approach to QRNG data processing as the following 5 step algorithm:
\begin{enumerate}
\item Run QRNG and collect $M$ independent samples $d = \{x_{1},\cdots,x_{M}\}$ from the distribution $f_{X}(x|\theta)$ defined by the QRNG.
\item Construct a prior $\pi(\theta)$ for all possible values of $\theta$.
\item Update the prior $M$ times using the Bayes rule Eq.(\ref{Eq:BayesRule}). Compute the posterior $\pi(\theta|d)$.
\item For each measurement outcome $x_{i}$ compute the correspondent distribution $g_{U}(u|x_{i})$ using Eq.(\ref{Eq:GofU}) and Eq.(\ref{EQ:IntTrans}). Set the acceptance probability value $P_a$
\item Use the proposed acceptance/rejection test to convert the measured sequence $d$ into integer values.  
\end{enumerate} 

It is worth mentioning that alternatively, instead of waiting to collect a measurement record $d$, one could choose to update the prior on-line i.e. after each measurement. In this case it is likely that a few first measurement results will be discarded as we accumulate information about the QRNG device at hand. However, after enough information is received to narrow down the parameter distribution, it will be possible to convert upcoming measurements into random bit values.  

\section{Examples}
To illustrate how our approach works in an experiment we consider two physical implementations of QRNGs. We first introduce mathematical models to describe the QRNGs of interest in the Section~\ref{Sec:PhysModels} and then proceed with the analysis of the experimental data and numerical simulations results in Section~\ref{Sec:RandD}. 
\subsection{Physical Models of QRNGs}\label{Sec:PhysModels}
\subsubsection{Photon Time-of-Arrival QRNG} \label{Sec:PhysModelsToA}
Let us first consider a QRNG based on measuring time-of-arrival statistics of a coherent light source. Our experimental setup consists of a tapered amplifier, emitting spontaneously and subsequently attenuated to a coherent state, that continuously illuminates the surface of a free-running single photon counting module with 80~ns dead time \cite{pooser}. Using a gated FPGA essentially acting as a time to digital converter, we measure the time interval $\tau$ between two consecutive photodetection events. The time interval $\tau$ plays the role of a physical random variable $X$ that we would like to convert into a discrete uniform random variable.  
  
To determine statistical properties of  $\tau$, a quantum model of phododetection process is needed. For this purpose we introduce a positive-operator valued measure (POVM)  $\{ \hat{{\rm P}}_{0}, \hat{{\rm P}}_{click}\} $, where $\hat{{\rm P}}_{0} = |0\rangle\langle 0|$ is a projection operator that corresponds to ``no-click'' measurement of the detector and  $\hat{{\rm P}}_{click}=\sum\nolimits_{k=1}^{\infty}|k\rangle\langle k|$ represents a ``click'' detection event. Note that $\hat{{\rm P}}_{click} + \hat{{\rm P}}_{0} = 1\!\!1$. Then the detector click rate (i.e the click probability per unit time) reads,
\begin{equation}
\frac{d P_{click}}{dt} = \theta {\rm Tr} [\rho_{l}\hat{{\rm P}}_{click}],
\end{equation}  
here $\rho_{l}$ is the density operator of the laser field and $\theta$ describes the overall detection efficiency. Therefore, the probability to get a click in a short time interval $\delta t$  is $P_{click} = \theta\delta t {\rm Tr} [\rho_{l}\hat{{\rm P}}_{click}]$. On the other hand, the 
probability to detect no click in the same time interval is $P_{0} = 1 - P_{click}$. Next consider the time interval $\tau$ between two consecutive detector clicks. We can model the absence of clicks during time $\tau$ by a sequence of $N$ successful ``no-click'' measurements each of the duration $\delta t = \tau/N$. Hence, the probability to observe no clicks during time $\tau$ reads,
\begin{equation}
P_{0}(\tau) = \left( 1-\frac{\tau \tilde{\theta}}{N} \right)^{N},
\end{equation}
where we introduced $\tilde{\theta} = \theta {\rm Tr} [\rho_{l}\hat{{\rm P}}_{click}]$. In the limit of large $N$ we obtain,
\begin{equation}
P_{0}(\tau) =\lim\limits_{ N \to \infty}\left( 1-\frac{\tau \tilde{\theta}}{N} \right)^{N} = e^{-\tilde{\theta}\tau}.
\end{equation}
We now can compute the conditional probability to detect a click at $t=\tau$ given a click was detected at $t=0$,
\begin{equation}\label{Eq:ProbTau}
P(\tau|0) = \frac{P_{click}(t=\tau)P_{0}(\tau)P_{click}(t=0)}{P_{click}(t=0)} = \tilde{\theta}\delta t  e^{-\tilde{\theta}\tau}.
\end{equation}
Finally, the probability density $f(\tau|\tilde{\theta})$ for the random variable $\tau$ can be obtained by taking a derivative of $P(\tau|0)$,
\begin{equation}\label{Eq:ToApdf}
f(\tau|\tilde{\theta}) = \frac{d P(\tau|0)}{d t} = \tilde{\theta} e^{-\tilde{\theta}\tau}.
\end{equation}
Two main assumptions were made in the derivation of Eq.(\ref{Eq:ToApdf}). First, the detection events are independent and identically distributed. This assumption is justifiable in case of moderate laser powers.
Second, we have assumed noiseless detection. The later assumption is, unfortunately, not very realistic. 

Avalanche photodiode detectors usually introduce two main sources of noise that affect the value of $\tau$ -- afterpulsing and timing jitter. Afterpulsing is a false detection event in which electrons that were trapped by quenching in a previous detector gate are rereleased in subsequent detector gates, usually occuring after a true click due to a photon absorption event. The time interval $\tau_{a}$ between a true detection and an afterpulse event can be well characterized experimentally and the raw data can be filtered to remove the afterpulsing events by only accepting measurements with $\tau\ge\tau_{a}$. The filtering procedure effectively results in rescaling of the probability density in Eq.(\ref{Eq:ToApdf}),
\begin{equation}\label{Eq:ToApdfRenorm}
f(\tau|\tilde{\theta}) = \tilde{\theta} e^{-\tilde{\theta}(\tau-\tau_{a})},
\end{equation}    
where $\tau_{a}$ is a characteristic afterpulsing time. 

The time jitter is a small error in the measurement of $\tau$. The recorded time interval between two sequential clicks $\tau_{r}$ is a sum of two random variables $\tau_{r} = \tau + \tau_{j}$, where $\tau$ is the ``true'' time interval with pdf $
f(\tau|\tilde{\theta})$ given in Eq.(\ref{Eq:ToApdfRenorm}) and $\tau_{j}$ is a time jitter random $\mathcal{N}(0,\sigma_{j}^2)$ variable. One can show that the probability density for $\tau_{r}$ reads,
 \begin{align}\label{Eq:ToApdfNoise} \nonumber
 f(\tau_{r}|\sigma_{j},\tilde{\theta}) =\hspace{5.7 cm}  \\
  \tilde{\theta}e^{-\tilde{\theta}(\tau_{r}-\tau_{a})+\frac{\sigma_{j}^{2}\tilde{\theta}^{2}}{2}}\left[ {\rm erf}(\frac{\tau_{r}-\tau_{a}-\sigma_{j}^2 \tilde{\theta}}{\sqrt{2}\sigma_{j}}) - {\rm erf}(\frac{\tilde{\theta}\sigma_{j}}{\sqrt{2}}) \right]
\end{align}
where ${\rm erf}$ denotes the error function. Notice that if the time jitter is small ($\sigma_{j}\rightarrow 0$), Eq.(\ref{Eq:ToApdfNoise}) coincides with Eq.(\ref{Eq:ToApdfRenorm}). Since the observed time jitter is indeed small we will model the time-of-arrival QRNG using the probability density in Eq.(\ref{Eq:ToApdfRenorm}) with one parameter $\tilde{\theta}$.

Let us also discuss how to implement the QRNG data processing algorithm described earlier for this model. The model pdf is given in Eq.(\ref{Eq:ToApdfRenorm}). An obvious choice for the prior is a non-informative (uniform) prior $\pi(\tilde{\theta}) = const$  that assigns constant weight to all values of the parameter $\tilde{\theta}$. It turns out that in this case the posterior distribution $\pi(\tilde{\theta}|{\tau_{1},\cdots,\tau_{n}})$ after $n$ measurements can be calculated even analytically (instead of standard numerical updating) as,
\begin{equation}\label{Eq:PosteriorToA}
\pi(\tilde{\theta}|{\tau_{1},\cdots,\tau_{n}}) = \Gamma(n+1,T) = \frac{\tilde{\theta}^n e^{-\frac{\tilde{\theta}}{T}}}{T^{n+1}n!},
\end{equation}
where $T=\left(\sum\limits_{k=1}^{n}\tau_{k}-n\tau_{a}\right)^{-1}$, and we assume that the characteristic afterpulsing time $\tau_{a}$ is known (not a parameter) and $\Gamma(n+1,T)$ denotes the gamma distribution function. Using Eq.(\ref{EQ:IntTrans}) we introduce $n$ random variables $u(\tilde{\theta}|\tau_{1}),\cdots,u(\tilde{\theta}|\tau_{n})$, where $u(\tilde{\theta}|\tau_{i}) = 1 - e^{-\tilde{\theta}(\tau_{i}-\tau_{a})}$ and compute their probability distribution $g_{i}(u_{i}|\tau_{i})$ using Eq.(\ref{Eq:GofU}) and Eq.(\ref{Eq:PosteriorToA}),
\begin{equation}
g_{i}(u_{i}|\tau_{i}) = \frac{[-\ln(1-u_{i})]^{n}(1-u_{i})^{\frac{1}{T(\tau_{i}-\tau_{a})}-1}}{n!(T(\tau_{i}-\tau_{a}))^{n+1}}.
\end{equation}
And finally we calculate the probability $P(u_{i}\in j) = P(\frac{j-1}{N}\le u_{i}\le \frac{j}{N} )$ that $u_{i}$ falls into the $j$-th bin ($j\in[1,N]$)
\begin{align}\label{Eq:ProbBinToA}\nonumber
P(u_{i}\in j) = \frac{1}{n!}[\gamma(n+1,-\frac{\ln(1-j/N)}{T(\tau_{i}-\tau_{a})}) - \\ 
 - \gamma(n+1,-\frac{\ln(1-(j-1)/N)}{T(\tau_{i}-\tau_{a})})],
\end{align}
where $\gamma$ denotes the lower incomplete gamma function. Applying the acceptance/rejection test to $P(u_{i}\in j)$ for all pairs $(i,j)$ will convert measurement outcomes $\tau_{1},\cdots,\tau_{n}$ into a sequence of uniformly distributed integers on $[1,N]$.
\subsubsection{Vacuum Quadrature Measurement QRNG}\label{Sec:HomodyneVacTheory} 
The second system that we consider here is a popular QRNG implementation based on vacuum quadrature measurement. Quantum vacuum fluctuations of the electromagnetic field are measured routinely at optical wavelengths using homodyne detection techniques~\cite{pironio}. A typical homodyne detector consists of a beam splitter with two input ($I_{1},I_{2}$) and two output ports ($O_{1},O_{2}$). Suppose that the input port $I_{1}$ carries a laser field described by a density operator $\rho_L$ and the port $I_{2}$ carries the vacuum. By placing a photodetector in each of the output ports we measure the photon number difference operator $\hat{N}_{-}$ between $O_{1}$ and $O_{2}$,
\begin{eqnarray}
\hat{N}_{-} & =  & [(\eta_{1}t)^2 - (\eta_{2}r)^2]a^{\dagger}a + [(\eta_{1}r)^2 - (\eta_{2}t)^2]b^{\dagger}b + \nonumber \\
&  & rt(\eta_{1}^2 + \eta_{2}^2)[a^{\dagger}b + ab^{\dagger}],
\end{eqnarray}
where $t$ ($r$) are the transmittance (reflectance) of the beam splitter, $\eta_{1,2}$ are detector 1,2 detection efficiencies and $a^{\dagger},a$ ($b^{\dagger},b$) are creation/annihilation operators for the input port $I_{1}$ ($I_{2}$). Therefore, in a general experimental situation, $\hat{N}_{-}$ will depend on three parameters $r, \eta_{1}, \eta_{2}$ (note that $t^2 = 1 - r^2$) and the laser field $\rho_L$. But since we only perform a numerical simulation of an experiment here and thus can ``control'' the parameters perfectly, we will assume that we have a 50/50 beam splitter ($t^2 = r^2 = 0.5$) and 100 percent efficient detectors ($\eta_{1}=\eta_{2}=1$).  We will also assume that the laser field is in a coherent state, i.e. $\rho_L = |\alpha\rangle\langle\alpha|$. Therefore, the expectation value of $\hat{N}_{-}$,
\begin{equation}
\langle \hat{N}_{-} \rangle =  |\alpha|\langle 0|b e ^{-i\phi_{\alpha}}+ b^{\dagger}e ^{i\phi_{\alpha}}|0\rangle = |\alpha|\langle 0|\hat{X}(\phi_{\alpha})|0\rangle,
\end{equation} 
is proportional to the expectation value of the vacuum quadrature operator $\hat{X}(\phi_{\alpha})$. Setting $|\alpha| = 1$ we conclude that by measuring the photon number difference in the output ports $O_{1}$ and $O_{2}$ we effectively measure the vacuum $X$ quadrature, and hence, a particular measurement outcome in a normal random variable $x_{vac}$ with pdf $\mathcal{N}(0,\sigma_{vac}^2)$. 

In reality measurement results are always affected by electronic noise. The noise is usually model by a normal distribution $\mathcal{N}(0,\sigma_{e}^2)$ and thus the outcome of the quadrature measurement  is a sum of the ``true'' quadrature random variable and the noise i.e. $x_{r} = x_{vac}+x_{e}$. Since $x_{vac}$ and $x_{e}$ are independent and normally distributed, their sum $x_{r}$ is also a normally distributed random variable with pdf $\mathcal{N}(0,\sigma_{vac}^2+\sigma_{e}^2)$. Therefore, we will model the output of vacuum quadrature measurement based QRNG as a continuous random variable $x_{r}$ with the distribution function $f(x|\sigma)$,
\begin{equation}\label{Eq:HQRNGpdf}
f(x|\sigma) = \frac{1}{\sqrt{2\pi}\sigma}e^{-\frac{x^2}{2\sigma^2}},
\end{equation} 
where $\sigma^2 = \sigma_{vac}^2+\sigma_{e}^2$ is an unknown parameter.

With the QRNG model at hand we can now discuss how to apply the data processing algorithm developed in Section~\ref{Sec:BayesianInference} to the vacuum quadrature measurement QRNG. Once again we start by choosing a prior. We propose to use a non-informative prior $\pi(\sigma) = const$ as in the previous example. The posterior distribution $\pi(\sigma|{x_{1},\cdots,x_{n}})$ after $n$ measurements can then be calculated analytically and reads,
\begin{equation}\label{Eq:PosteriorHQRNG}
\pi(\sigma|{x_{1},\cdots,x_{n}}) = \frac{X^{\frac{n-1}{2}}e^{-\frac{X}{2\sigma^2}}}{\sqrt{2^{n-3}}\Gamma(\frac{n-1}{2})\sigma^{n}},
\end{equation} 
where $X=\sum\limits_{i=1}^{n}x_{i}^2$ and $\Gamma(\frac{n-1}{2})$ denotes the gamma function.

The next step in our procedure is to introduce $n$ random variables $u(\sigma|x_{1}),\cdots,u(\sigma|x_{n})$ that later on will be binned. Unlike the previous example, where Eq.(\ref{EQ:IntTrans}) was used for that purpose, we will rely on {\it Box-Muller} transform~\cite{CasellaBerger} here. Recall that $U_{1}$ and $U_{2}$, two independent uniform(0,1) random variables, can be converted into two independent normal $\mathcal{N}(0,1)$ random variable $X$ and $Y$ using the following transformation,
\begin{eqnarray}
X = R\cos{\theta} & &\hspace{2cm} R = \sqrt{-2\ln{U_{1}}}  \nonumber \\
Y = R\sin{\theta} & &\hspace{2cm} \theta = 2\pi U_{2},
\end{eqnarray}
On the other hand, a pair of measurement outcomes $x_{i1}/\sigma$ and $x_{i2}/\sigma$ can be converted into two random variables $u_{1}(\sigma|x_{i1},x_{i2}) = \exp(-\frac{x_{i1}^2+x_{i2}^2}{2\sigma})$ and $u_{2}(\sigma|x_{i1},x_{i2}) = \arctan(x_{i2}/x_{i1})/2\pi$ $\in$ [0,1]. Since $u_{2}(\sigma|x_{i1},x_{i2})$ does not depend on the parameter $\sigma$ (it is constant for a given pair $x_{i1},x_{i2}$ ), it can be immediately placed into the $j$-th bin that satisfies $(j-1)/N\le u_{2} \le j/N$. As to $u_{1}(\sigma|x_{i1},x_{i2})$ which indeed is a function of $\sigma$, we can derive its probability distribution function $g(u_{1}|x_{i1},x_{i2})$ using the posterior distribution in Eq.(\ref{Eq:PosteriorHQRNG}),
\begin{equation}
g(u_{1}|x_{i1},x_{i2}) = \left(\frac{X}{x_{i1}^2+x_{i2}^2}\right)^{\frac{n-1}{2}}\frac{(-\ln{u_{1}})^{\frac{n-3}{2}}u_{1}^{\frac{X}{x_{i1}^2+x_{i2}^2}-1}}{\Gamma(\frac{n-1}{2})}.
\end{equation}
Finally,  the probability $P(u_{1}\in j) = P(\frac{j-1}{N}\le u_{1}\le \frac{j}{N} )$ that $u_{1}$ falls into the $j$-th bin ($j\in[1,N]$)
\begin{align}\label{Eq:ProbBinHQRNG}\nonumber
P(u_{1}\in j) & = \frac{1}{\Gamma(\frac{n-1}{2})}[\gamma(\frac{n-1}{2},-\frac{X\ln(\frac{j-1}{N})}{x_{i1}^2+x_{i2}^2}) - \\ 
& - \gamma(\frac{n-1}{2},-\frac{X\ln(\frac{j}{N})}{x_{i1}^2+x_{i2}^2})],
\end{align}
where $\gamma$ is the lower incomplete gamma function. Applying the acceptance/rejection test to $P(u_{1}\in j)$ for all $j$ will convert $u_{1}$ into a uniformly distributed integer on $[1,N]$. Therefore, a pair of normally distributed outputs of the vacuum homodyne measurement $x_{i1}$ and $x_{i2}$ converts into two uniformly distributed integer random numbers $j_1$ and $j_2$. 
\begin{figure}[t]
	\begin{center}
		\includegraphics[scale = 0.4]{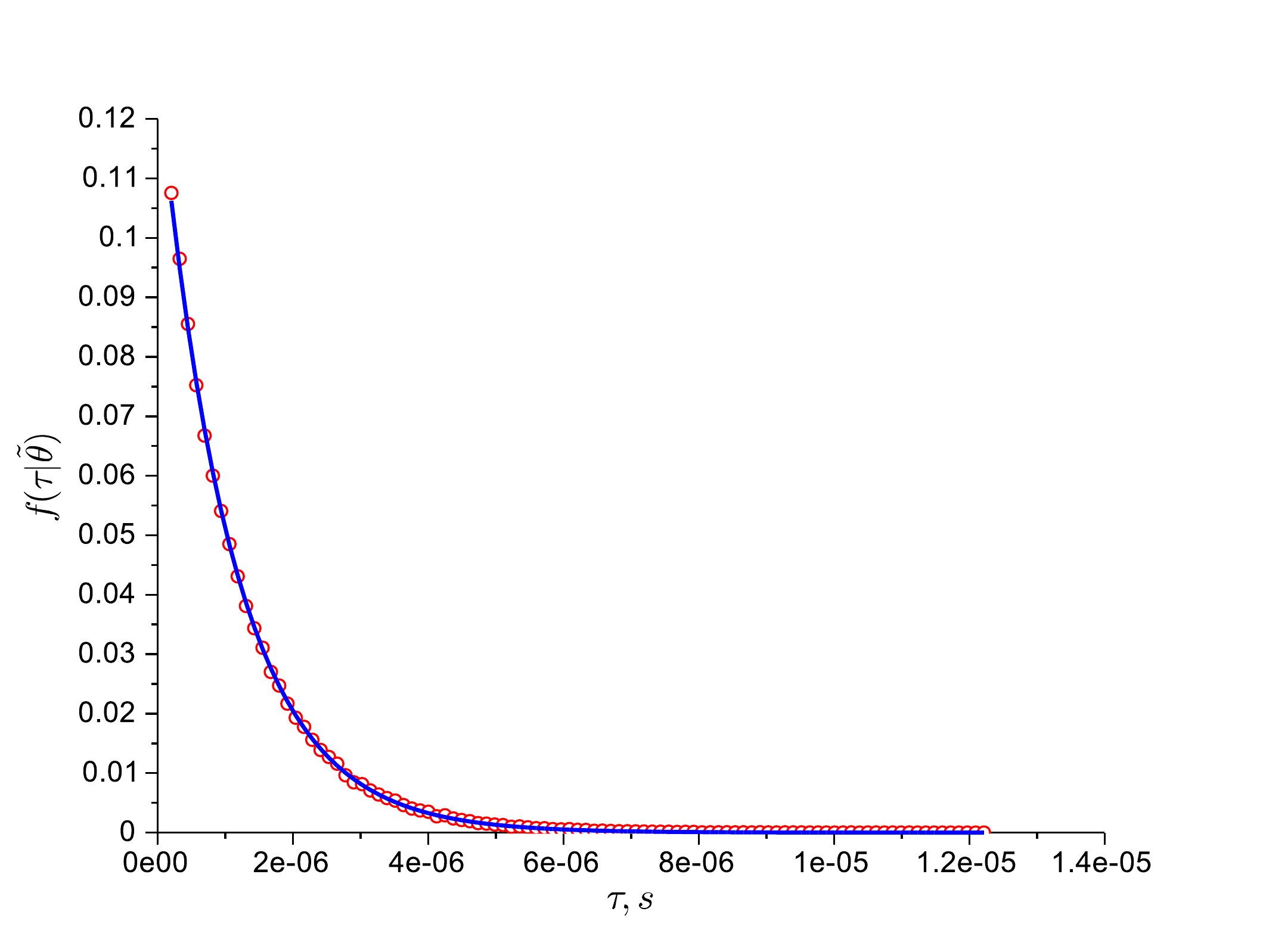}
		\caption{Probability distribution of the time intervals between two successive detection events observed in the experiment.}
		\label{Fig:ToAdataPdf}
	\end{center}
\end{figure}

\subsection{Experimental Results and Simulations}\label{Sec:RandD}
\subsubsection{Photon Time-of-Arrival QRNG}

We collected a sample containing 256,000 measurements of the time interval between two consecutive detection events \cite{pooser}. The raw data was filtered and all entries $\tau<\tau_{a} = 7.81\times10^{-8} s$ were removed from the sample to mitigate the effect of detector afterpulsing. The resulting filtered sample consisted of 221,890 measurements. We binned the filtered data into 100 bins of equal size $\Delta\tau = 1.225\times10^{-7}s$ and calculated the probability of each bin. The corresponding probability distribution is depicted in Fig.~\ref{Fig:ToAdataPdf} with red circles. Based on the QRNG model discussed in Section~\ref{Sec:PhysModelsToA} we calculated $\tilde{\theta}_{ML} = 9.16\times10^5s^{-1}$, MLE for the parameter $\tilde{\theta}$. We used $\tilde{\theta}_{ML}$ in conjunction with the probability density function in Eq.(\ref{Eq:ToApdf}) to fit the experimental data. The result is depicted on Fig.~\ref{Fig:ToAdataPdf} with the solid blue line. Not surprisingly, given the number of measurements, the ML curve fits the data well. 
\begin{figure}[t]
	\begin{center}
		\includegraphics[scale = 0.4]{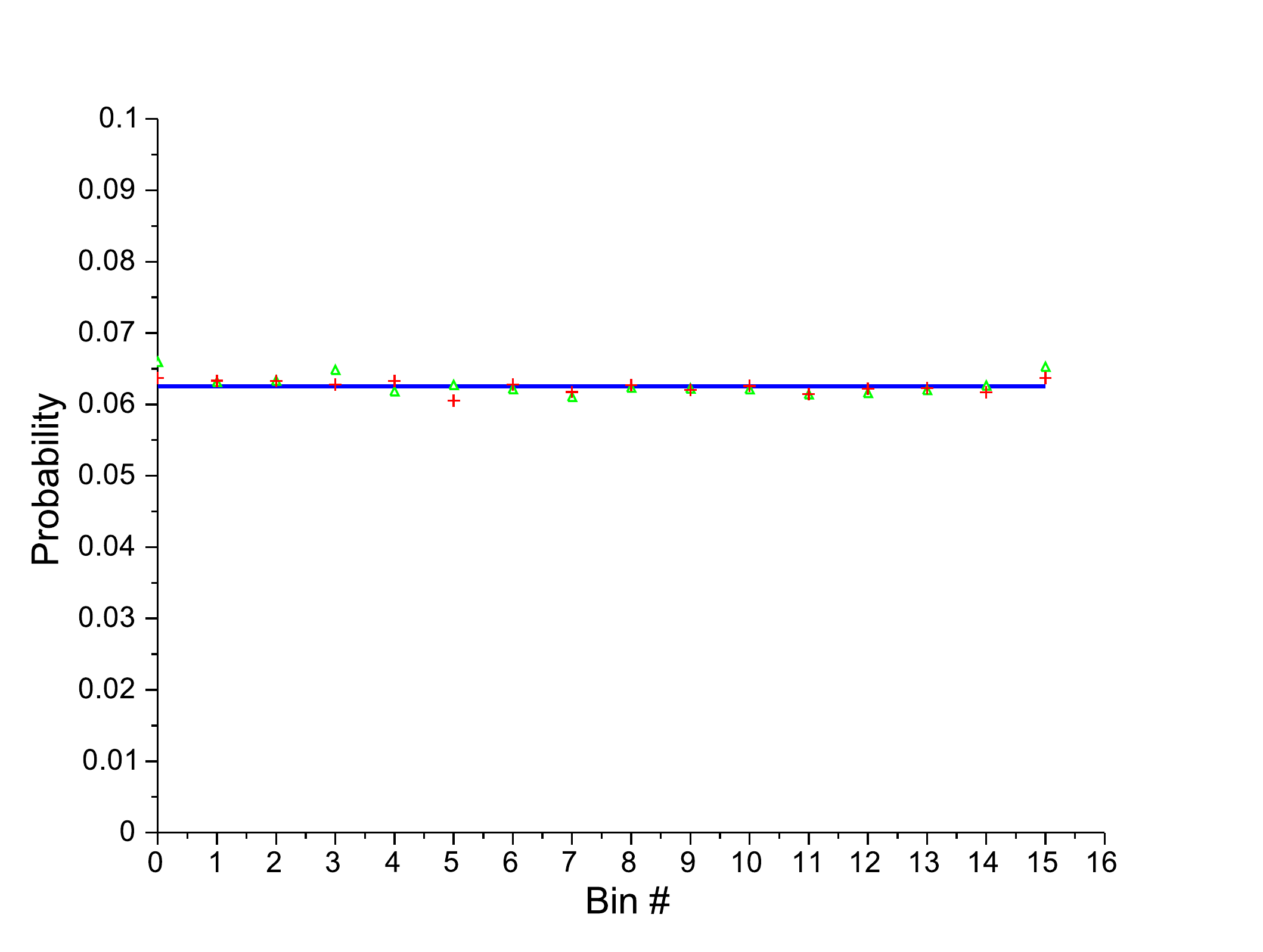}
		\caption{Probability distribution of the 4-bit random numbers.}
		\label{Fig:ToA4bitPdf}
	\end{center}
\end{figure}
\begin{figure}[!t]
	\begin{center}
		\includegraphics[scale = 0.4]{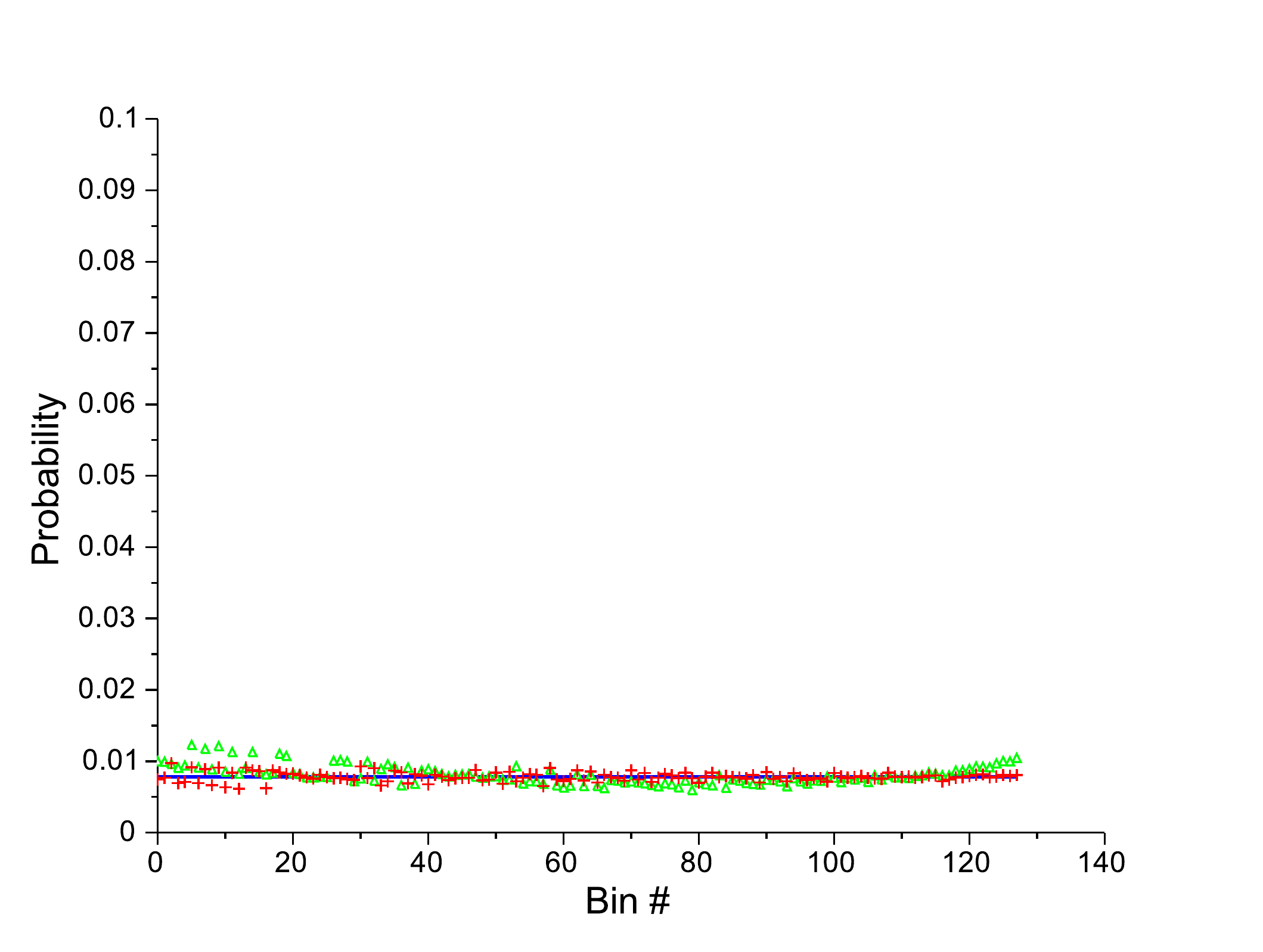}
		\caption{Probability distribution of the 7-bit random numbers.}
		\label{Fig:ToA7bitPdf}
	\end{center}
\end{figure}
\begin{figure}[!t]
	\begin{center}
		\includegraphics[scale = 0.4]{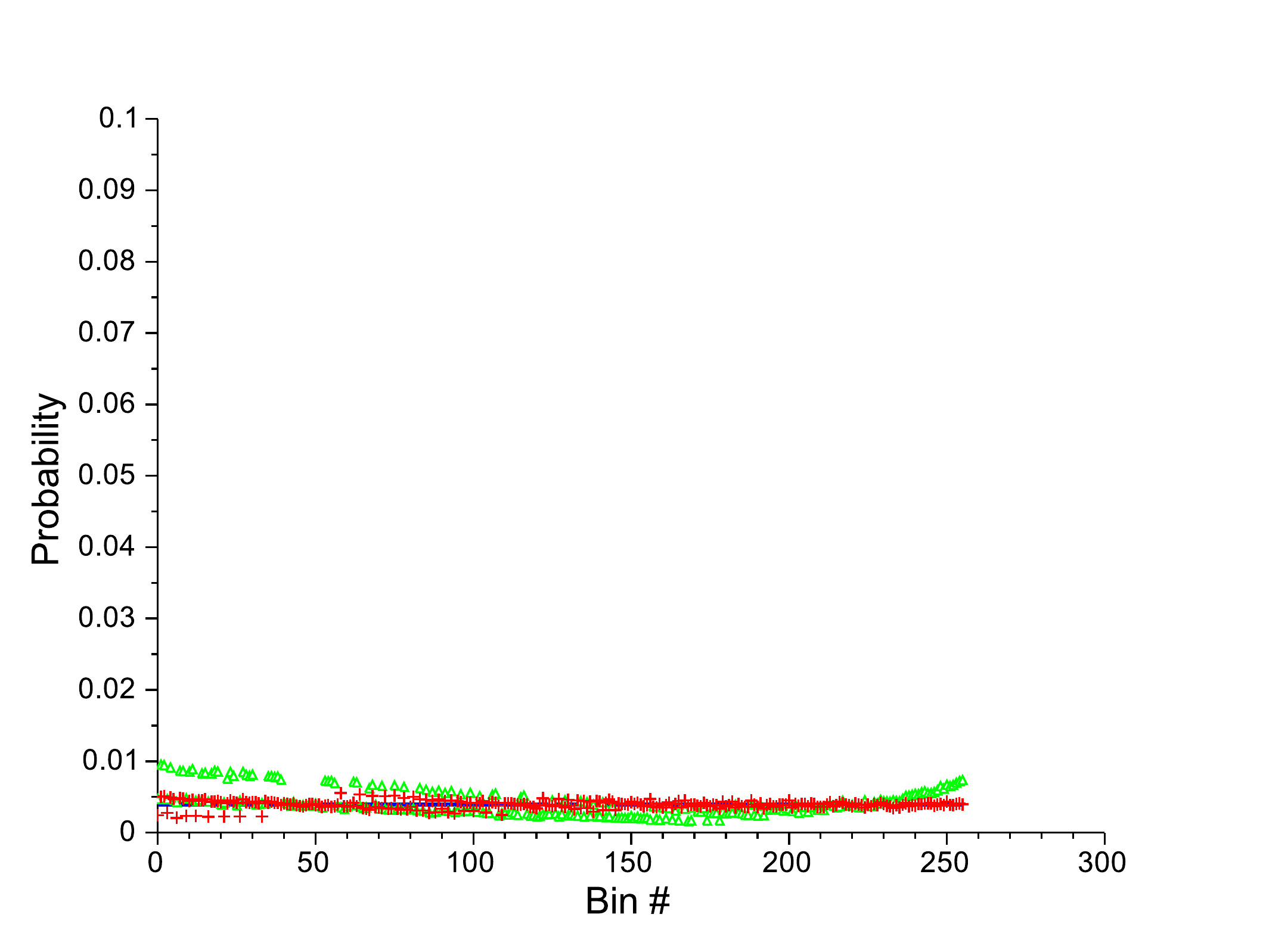}
		\caption{Probability distribution of the 8-bit random numbers.}
		\label{Fig:ToA8bitPdf}
	\end{center}
\end{figure}

Next we applied our data processing algorithm to the filtered data. We set the acceptance probability $P_{a}=0.95$ and proceeded to convert the data into a set of 4-bit random numbers (i.e. measurement results are binned among $2^{4}=16$ bins). The number of measurements that passed acceptance/rejection criterion ($P\ge P_{a}$), and were assigned a bin value ($0,1,\cdots,15$), was 215,538 (out of 221,890). The resulting bin probability distribution is depicted in Fig.~\ref{Fig:ToA4bitPdf} using green triangles. The solid blue line corresponds to the ideal 4-bit uniform distribution and the red crosses represent a 4-bit probability distribution obtained from the same data set using the conventional fixed-parameter binning technique with $\tilde{\theta} = \tilde{\theta}_{ML}$. Both methods generate a visually uniform distribution. The uniformity is also confirmed by the values of Shannon entropy  per bit -- $H$ -- for each distribution. For the conventional binning $H = 0.999966$ bits and the entropy of the distribution generated by our binning method is $H = 0.999914$ bits.

In conventional bin assignment methods, once the distribution parameter value is estimated from a given set of measurements, the number of random bits that can be generated per single measurement is, in principle, only limited by the number of measurements~\cite{NumericalNoise}. This is because the mean error (standard deviation) of the parameter estimator is ignored in conventional binning. However, if the parameter estimation error is greater than the width of the bin where the measurement result is placed then such a bin assignment is erroneous and this measurement must be ignored and removed from the data. But this is exactly what our bin assignment method with the acceptance probability $P_{a}=0.95$ does. It effectively requires that the bin width should be greater than 4 standard deviations of the random variable $u_{i}$. If this requirement is not fulfilled the $i$-th measurement can not be assigned a bin reliably and the measurement is discarded. Hence, in contrast to the conventional binning our approach reduces the overall number of measurements. Therefore, for a given initial set of data, the number of random bits per measurement is naturally less in our method. In other words Bayesian updating provides a more conservative estimate of randomness of a QRNG when compared to ad-hoc binning. To illustrate this we generated 7- and 8-bit random number distributions from the same filtered data that we used for the 4-bit distribution above and the acceptance probability $P_{a}=0.95$. The resulting distributions are depicted with green triangles on Fig.~\ref{Fig:ToA7bitPdf} and Fig.~\ref{Fig:ToA8bitPdf}. As before the solid blue line corresponds to the ideal 7(8)-bit uniform distributions and the red crosses represent 7(8)-bit probability distribution obtained from the same size data sets using the conventional fixed-parameter binning technique with $\tilde{\theta} = \tilde{\theta}_{ML}$. The number of measurements that have passed acceptance/rejection criteria ($P\ge P_{a}$) is 172,736 (122,927) for the 7-(8-)bit distribution. We also calculated Shannon entropy for the conventional binning, $H_{7bit} = 7\times0.999314$ bits, and the entropy of our binning method is $H_{7bit} = 7\times 0.997237$ bits. On the other hand the entropy in the 8-bit case for conventional binning is $H_{8bit} = 8\times 0.998070$ bits and for the proposed binning method is $H_{8bit} = 8\times 0.981067$ bits. As previously suspected, we observe a drop in the entropy of the 8-bit distribution generated using our technique. This implies that the collected data can reliably be converted into random bit sequences up to 7 bits long. Note that the conventional binning method does not provide us with such a conclusion.

\subsubsection{Vacuum Quadrature Measurement QRNG} 
\begin{figure}[!t]
	\begin{center}
		\includegraphics[scale = 0.4]{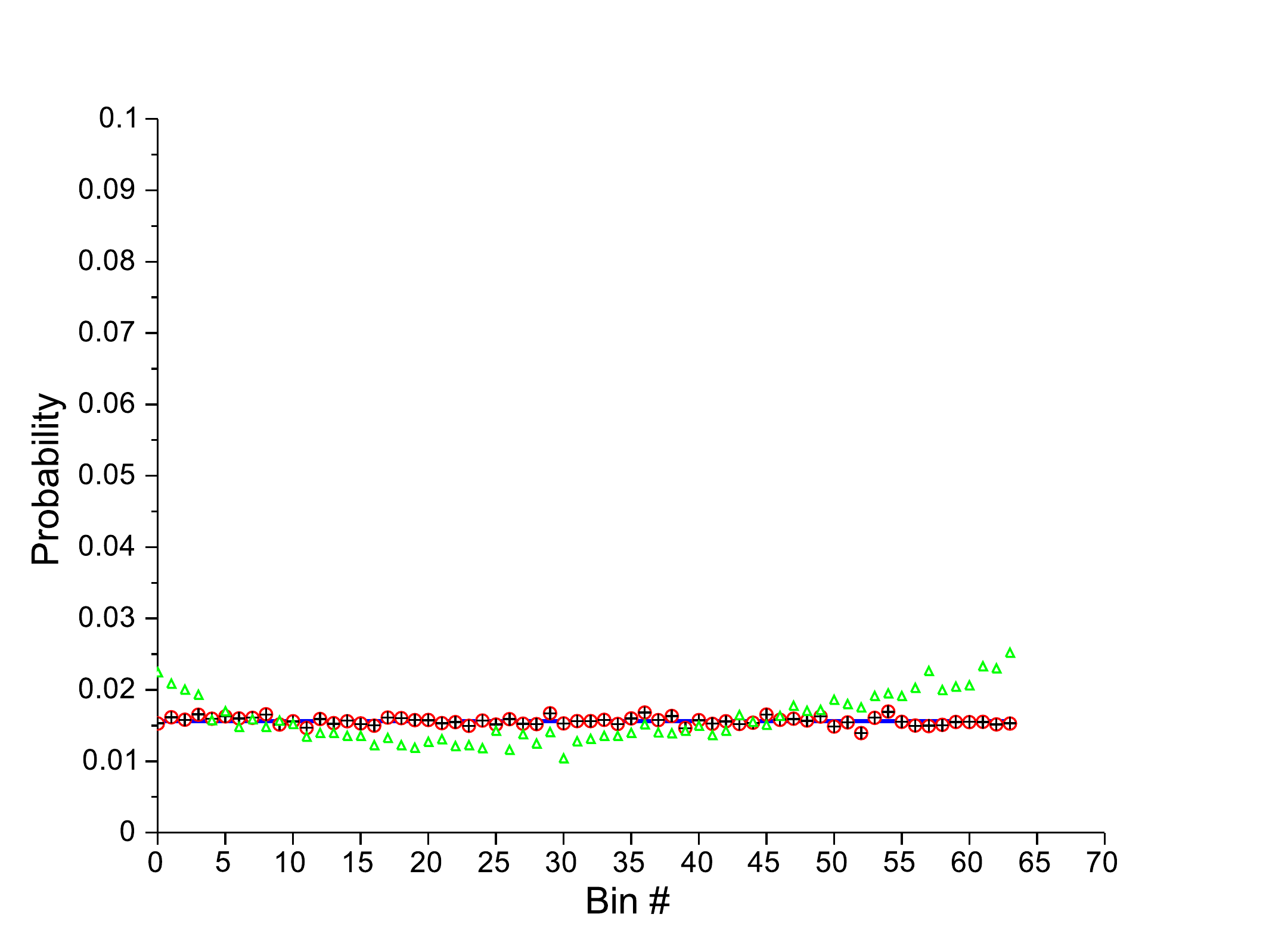}
		\caption{Probability distribution of the 6-bit random numbers.}
		\label{Fig:VacHomodyne6bitPdf}
	\end{center}
\end{figure}

We simulated vacuum homodyne measurements using a pseudo random number generator. Two independent sets of 50,000 random numbers were created by sampling the normal distributions $\mathcal{N}(0,\sigma_{vac})$ and $\mathcal{N}(0,\sigma_{e})$ respectively. The first set with $\sigma_{vac}=1$ represents noiseless vacuum quadrature measurement whereas the second set with $\sigma_{e}=0.1\sigma_{vac}$ corresponds to the electronics noise. Thus, the sum of the sets simulates the vacuum homodyning based QRNG that we previously modeled using Eq.(\ref{Eq:HQRNGpdf}). 
\begin{figure}[!t]
	\begin{center}
		\includegraphics[scale = 0.4]{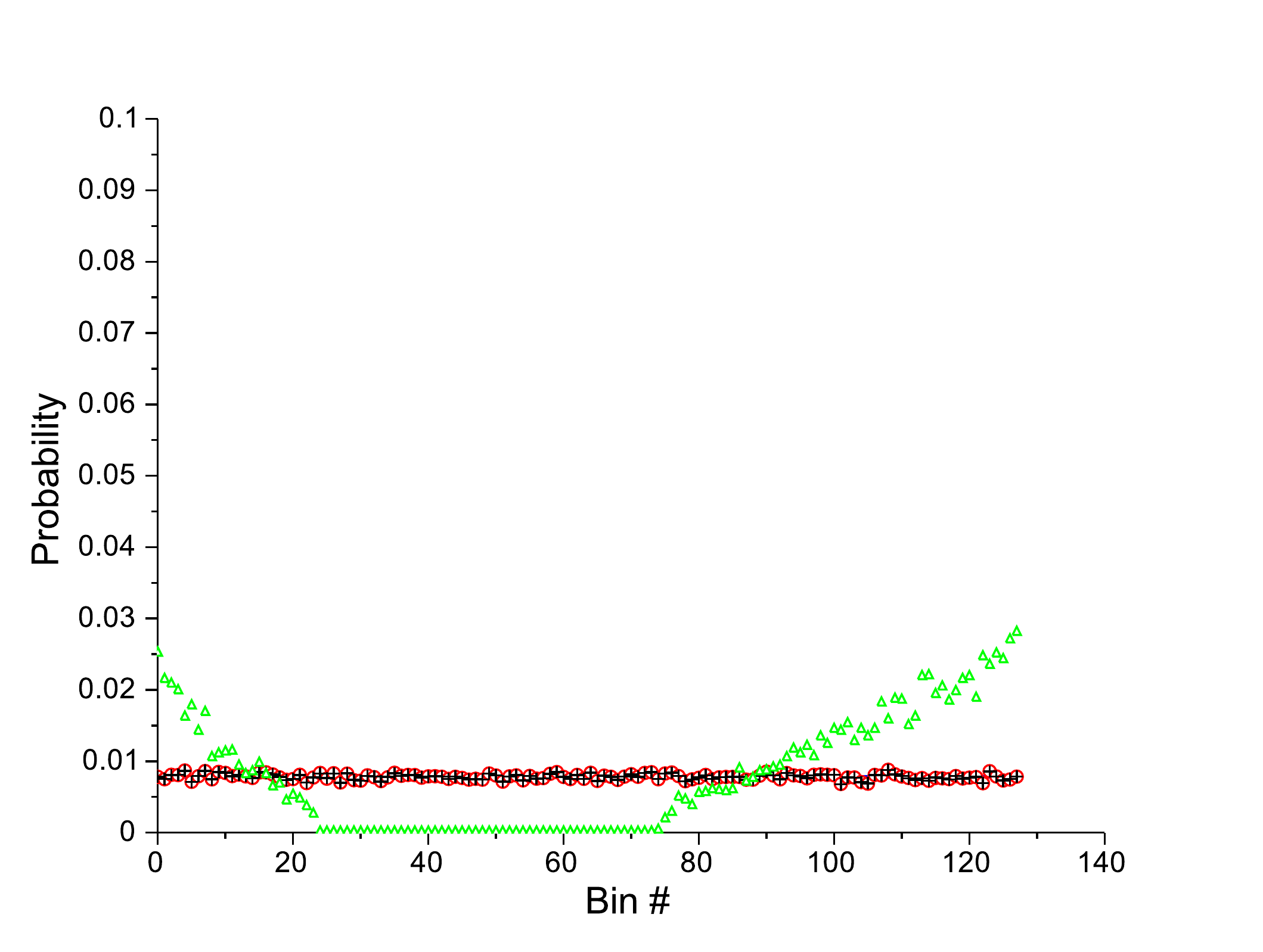}
		\caption{Probability distribution of the 7-bit random numbers.}
		\label{Fig:VacHomodyne7bitPdf}
	\end{center}
\end{figure}

We used the data to produce sets of 6- and 7-bit random numbers implementing both the conventional (MLE based) and proposed (Bayesian) binning methods. The resulting distributions are depicted in Fig.~\ref{Fig:VacHomodyne6bitPdf} and Fig.~\ref{Fig:VacHomodyne7bitPdf}. The green triangles correspond to the probability distributions generated using our technique ($P_{a}=0.95$), the red circles depict the results of the conventional MLE based binning and black crosses represent conventional binning with the ``true'' value of the parameter $\sigma^2 = 1.1$. 

Examining Fig.~\ref{Fig:VacHomodyne6bitPdf} and Fig.~\ref{Fig:VacHomodyne7bitPdf} visually we observe that our method fails to produce a uniform 7-bit distribution indicating that the maximum number of random bits per measurement outcome cannot exceed 6 for the simulated data sample. This is also confirmed by the values of Shannon entropy $H_{6bit}=6\times0.9945876$ versus $H_{7bit}=7\times0.8668848$. Of course, generating a larger sample of measurements would allow a higher number of bits per measurement outcome as was the case in the previous Section. This illustrates the interplay between the number of measurement in a sample, acceptance probability, and the number of random bits that can be extracted from the sample.

\section{Summary}
In this manuscript we have demonstrated a new binning technique for QRNGs, as well as a formalized approach to characterize traditional binning methods. In particular, ad-hoc binning approaches are shown to result in possible bias when the model of the physical QRNG system is not taken into account. Using Bayesian hypothesis updating, a physical model can be used to quickly characterize experimental data. This has implications for new types of quantum statistical tests for randomness in a potentially more accessible manner than loop-hole-free Bell Inequality violation tests.

\begin{acknowledgements}
 P. L. would like to thank Bing Qi, Ryan Bennink and Travis Humble for useful discussions. 
This work was performed at Oak Ridge National Laboratory, operated by UT-Battelle for the U.S. Department of energy under contract no. DE-AC05-00OR22725.
\end{acknowledgements}

\end{document}